\documentclass[]{article}

\usepackage{epsfig,graphicx,float,pst-all}
\usepackage{amssymb}\usepackage{bbm}
\DeclareMathAlphabet{\mathpzc}{OT1}{pzc}{m}{it}

\newcommand{\be}{\begin{equation}} 
\newcommand{\ee}{\end{equation}}
\newcommand{\bea}{\begin{eqnarray}} 
\newcommand{\eea}{\end{eqnarray}}
\newcommand{\bc}{\begin{center}}
\newcommand{\ec}{\end{center}}
\newcommand{\un}[1]{\underline{#1}}

\begin{document}

\title{Diagonalization scheme for the many-body Schr\"odinger equation}
\author{Lorenzo Fortunato\footnote{\tt fortunat@pd.infn.it}, Tomohiro Oishi\footnote{\tt toishi@pd.infn.it} \\
\small Dipartimento di Fisica e Astronomia ``G.Galilei'', \\
\small I.N.F.N., Sezione di Padova,\\
\small via Marzolo 8, I-35131 Padova, Italy 
}
\maketitle
\begin{abstract}
A new convenient method to diagonalize the non-relativistic many-body Schr\"o\-dinger equation with two-body central potentials is derived. It combines kinematic rotations (democracy transformations) and exact calculation of overlap integrals between bases with different sets of mass-scaled Jacobi coordinates, thereby allowing for a great simplification of this formidable problem. We validate our method by obtaining a perfect correspondence with the exactly solvable three-body ($N=3$) Calogero model in 1D. 
\end{abstract}

\section{Outline of the method}
A new method to diagonalize the non-relativistic many-body Schr\"odinger equation (\ref{se}) with two-body central potentials, $V_{ij}$ is described.
The program develops through the following main steps:
1) use of mass-scaled Jacobi coordinates to separate out the center of mass motion; 
2) expansion of the wavefunction in a totally decoupled basis (tensor product of known orthonormal complete bases, one for each internal coordinate) ; 
3) calculation of matrix elements in this basis: those of $T$ and $V_{12}$ are trivially calculated, while, for the calculation 
of matrix elements of $V_{ij}$, we propose the following steps:
\begin{itemize}
\item[i)]{ kinematic rotations (democracy transformations) to alternative sets of Jacobi coordinates in which ${ij}$ becomes the 'first' coordinate}
\item[ii)]{ analytic calculation of displaced overlap integrals ($\sigma$ coefficients)}
\end{itemize}
The last two points allow a great simplification in the treatment of this long-standing quantum problem, reducing essentially all computations to one-dimensional integrals. This method has the advantage of pushing the analytical treatment as far as possible, without introducing too many complications. Points 1) to 3) are part of standard procedures, but 3), that is the core of the new method, contains new ideas. In particular, when the calculations in ii) is obtained analytically, there is an advantage in our method. After this program is achieved in a general theory for $N$ particles in 3D, we will demonstrate its validity in the case of three bosons in 1D, by reproducing in a numerical code the well-known analytic results of Calogero \cite{Cal1,Cal2}. This opens the way to a large class of problems that can be easily studied with our formalism.

\subsection{Mass-scaled Jacobi coordinates}
The three dimensional Schr\"odinger equation for $N$ interacting particles with masses $m_i$, $i=1,\dots, N$ is:
\be
\Biggl[\sum_{i=1}^{N}\frac{\vec p_i\,^2}{2m_i} + \sum_{i<j}^{N}V_{ij}(\mid \vec r_{ij}\mid)-E_T\Biggr]\Psi^{(N)}(\vec r_1, \dots , \vec r_N) =0 \;,
\label{se}
\ee
where $\vec  r_{ij} = \vec r_j -\vec r_i $. The problem of separating the center of mass motion has been recently re-analyzed  in Ref. \cite{Fort} for equal masses, while here we are interested in the generalization to different masses.
Based on Refs.\cite{De-Aq}, we introduce $N-1$ Jacobi coordinates, $\vec \xi_k$, plus the position of 
the center of mass :
\be
\left\{ 
\begin{array}{ccc}
\vec \xi_k &=& \sqrt{\frac{\mu_k}{\mu}} \bigl( \vec r_{k+1} -\vec c_k \bigr) \\
\vec \xi_N &=& \vec c_N \equiv \vec R
\end{array}
\right.
\label{tr}
\ee
with $k=1,\dots, N-1$ and where the vectors $ \vec c_k= \frac{\sum^k m_i \vec r_i}{\sum^k m_i} $
have been used for the sake of simplicity. 
The $\vec \xi_k$ coordinates correspond to the relative position vectors connecting the center of mass of the $k-$th cluster (i.e. the cluster made up of the first $k$ particles, in some, previously adopted, ordering sequence) and the $(k+1)-$th particle. The quantities
\be
\mu_k=\frac{(\sum_{}^k m_i)m_{k+1}}{\sum_{}^{k+1} m_i} \qquad k=1,\dots, N-1
\ee
are intermediate reduced-masses between the cluster of the first $k$ particles and the single $(k+1)$-th particle, 
while $\mu=\sqrt[N-1]{\prod^{N} m_i /\sum^{N} m_i}$ is a common reduced-mass.
The momenta $\vec \pi_k$ ($k=1,\dots, N-1$) are canonically conjugate to $\vec \xi_k$, and $\vec \pi_N\equiv \vec P$ is the total momentum, canonically conjugate to the center of mass position. They are related to the lab frame momenta through
\be
\vec p_k = \sum_l \Bigl(\frac{\partial \vec \xi_l}{\partial \vec r_k} \Bigr)\vec \pi_l \;.
\ee
The transformation of coordinates (\ref{tr}) leaves invariant the quadratic forms corresponding to total moment of inertia and kinetic energy: 
\bea
\sum_{i=1}^N m_i {\vec r_i}\,^2 &=&  \mu \sum_{i=1}^{N-1} {\vec \xi_i}\,^2  +M\vec R\,^2 \\
\sum_{i=1}^N \frac{\vec p_i\,^2}{2m_i} &=& \sum_{i=1}^{N-1} \frac{\vec \pi_i^2}{2\mu} + \frac{\vec P^2}{2M} \;,
\label{2fo}
\eea
where $M=\sum_{}^N m_j$ is the total mass of the system.
The property (\ref{2fo}) allows the exact separation of the center of mass motion.
By factorizing $\Psi^{(N)}(\vec r_1, \dots , \vec r_N)=f(\vec R) \Phi^{(N)}(\vec \xi_1, \dots , \vec \xi_{N-1})$ with $f(\vec R)= e^{i\vec K \cdot \vec R}$ and using $E_T=E+E_{cm}$, with $E_{cm}=\hbar^2\vec K^2/2M$, one obtains the Schr\"odinger equation in the intrinsic coordinates:
\be
\Biggl[\sum_{i=1}^{N-1}\frac{\vec \pi_i^2}{2\mu} + \sum_{i<j}^{N}V_{ij}(r_{ij})-E\Biggr]\Phi^{(N)}(\vec \xi_1, \dots , \vec \xi_{N-1}) =0 \;.
\ee
Note that we have not changed the potential energy term and its argument as there will be no need for this complication.
It is useful to note that $\vec r_{12} = \sqrt{\mu_1/\mu}~\vec \xi_1$ are collinear proportional vectors.

\subsection{Trivial matrix elements}
\begin{figure}[!t]
\begin{center}
\psscalebox{0.75}{
\begin{picture}(280,160)
\psset{unit=1.pt}
\psframe[framearc=0.1](30,10)(250,150)
\psline[linestyle=dashed](140,10)(140,150)
\psline[linestyle=dashed](30,80)(250,80)
\psframe[framearc=0.2](60,100)(120,120) \rput(90,110){Harm.Osc.}
\psframe[framearc=0.2](160,100)(220,120) \rput(190,110){Coulomb}
\psframe[framearc=0.2](60,40)(120,60)   \rput(90,50){Davidson}
\psframe[framearc=0.2](160,40)(220,60)   \rput(190,50){Kratzer}
\psline[doubleline=true]{<->}(90,100)(90,60)
\psline[doubleline=true]{<->}(190,100)(190,60)
\psline{<->}(120,110)(160,110)
\psline{<->}(120,50)(160,50)
\rput(140,80){\psframebox*{SU(1,1)}}
\rput(140,25){\psframebox*{``Deformed''}}
\rput(140,135){\psframebox*{``Spherical''}}
\rput(45,80){\rotateleft{\psframebox*{Asymp. $\rightarrow \infty$}}}
\rput(235,80){\rotateleft{\psframebox*{Asymp. $\rightarrow 0$}}}
\rput(140,156){DOUBLE DUALITY SCHEME}
\end{picture}}\end{center}
\caption{Dualities between analytically solvable harmonic-type 
potentials (Harmonic Oscillator and Davidson potentials) and coulomb-type potentials 
(Coulomb and Kratzer potentials). The SO(2,1)$\sim$SU(1,1) group structure is common
to all cases. The two right quadrants correspond to potentials that 
asymptotically go to zero as $1/r$, while the left ones go to infinity 
as $r^2$ (therefore they do not admit unbound eigenstates). 
The two lower quadrants identify potentials with the minimum in some
non-null finite point, while the two upper ones identify 
``spherical'' cases (minimum in $r_0=0$ for the harmonic oscillator case 
and singularity for the Coulomb case). Simple arrows indicate 
homology in the algebraic treatment (same commutation relations, different 
representations in terms of differential operators), while double arrows 
indicate an extension of the commutation relations that allows a formal 
analogy between the deformed and spherical cases (namely $\hat Z_1\rightarrow 
\hat Z_1+k/\hat Z_2$, with notation of Refs. \cite{al-te}).}
\label{Dual}  
\end{figure}
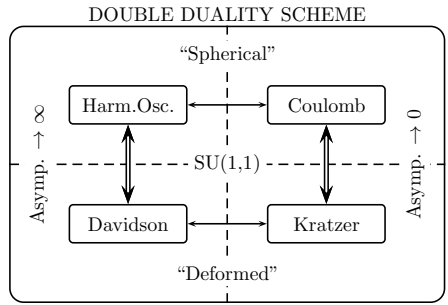

One has freedom to choose among the most suitable basis for his scopes: leaving aside the square well, there are essentially four types of potentials that one can choose from, each of which can be seen as a particular case of SU(1,1) dynamical symmetry. They are illustrated in Fig. \ref{Dual} with all the dualities that connect them. 

Totally decoupled basis states can be expressed as a tensor product of the $N-1$ relative motion wavefunctions that amounts to:
$$
\mid \tilde \Phi^{(N)}_{c} \rangle =\mid \phi^1_{\{\nu_1\}}(\vec \xi_1) \times \dots \times \phi^{N-1}_{\{\nu_{N-1}\}}(\vec \xi_{N-1}) \rangle $$
\be \rightarrow \mid \phi^1_{\{\nu_1\}}(\vec \xi_1) \dots \phi^{N-1}_{\{\nu_{N-1}\}}(\vec \xi_{N-1}) \rangle\;,
\label{expa} 
\ee
where each $\{\nu_i \}= \{ n_i, \ell_i,m_i \}$ indicates the set of all quantum numbers needed to specify each of the single motion wavefunctions (namely the principal q.n., the angular momentum q.n. and its projection on the quantization axis). The tensor product with good angular momentum is expressed through angular momentum coupling rules into sums of simple products, that can more practically be used as basis elements and one can forget the Clebsch-Gordan coefficients because these are reabsorbed into the amplitudes in the process of diagonalization. For ease of notation, in the following we will indicate each basis element with a single index, $c$,  that counts the basis states.

Within these basis states the matrix elements of $T_i$ are trivially calculated, because $\vec \pi_i^2$ acts only on the $i-$th coordinate, while all the remaining coordinates lead to Kronecker delta's in the respective quantum numbers:
\bea 
\label{met} 
\langle \tilde \Phi^{(N)}_{c'} \mid T_i \mid  \tilde \Phi^{(N)}_c \rangle =   \delta_{\{\nu_1'\}\{\nu_1\}} \dots \delta_{\{\nu_{i-1}'\}\{\nu_{i-1}\}}  \nonumber \\
\langle \phi^{i}_{\nu_i'}(\vec \xi_i) \mid T_i \mid  \phi^{i}_{\nu_i}(\vec \xi_i) \rangle\delta_{\{\nu_{i+1}'\}\{\nu_{i+1}\}} \dots \delta_{\{\nu_{N-1}'\}\{\nu_{N-1}\}}, 
\eea
and the same is true for the matrix elements of $V_{12}$ because its argument is already one of the Jacobi coordinates (the first):
\bea 
\label{meu} 
\langle \tilde \Phi^{(N)}_{c'} \mid V_{12} \mid  \tilde \Phi^{(N)}_c \rangle = \langle \phi^{1}_{\nu_1'}(\vec \xi_1) \mid V_{12}(\sqrt{\mu_1/\mu}\vec \xi_1) \mid  \phi^{1}_{\nu_1}(\vec \xi_1) \rangle \nonumber \\
\delta_{\{\nu_{2}'\}\{\nu_{2}\}} \dots \delta_{\{\nu_{N-1}'\}\{\nu_{N-1}\}} \;.
\eea

\subsection{Democracy transformation and non-trivial matrix elements}
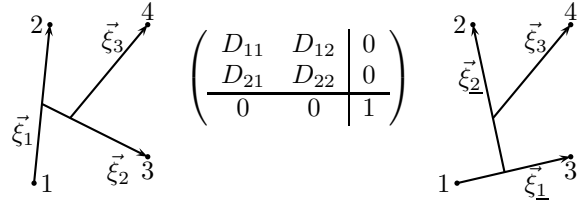
\begin{figure}[!t]
\begin{center}
\begin{picture}(205,80)
\psset{unit=1pt}
\psline{->}(0,0)(6,60) \pscircle*(0,0){1} \rput(5,0){1} \pscircle*(6,60){1} \rput(1,60){2}
\psline{->}(3,30)(43,10) \pscircle*(43,10){1} \rput(43,5){3}
\psline{->}(13.5,24.5)(43,60)\pscircle*(43,60){1} \rput(43,65){4}
\rput(-4,20){$\vec \xi_1$}
\rput(32,5){$\vec \xi_2$}
\rput(30,55){$\vec \xi_3$}
\rput(100,40){$\left( \begin{array}{cc|c} 
D_{11} &D_{12}& 0 \\ 
D_{21} &D_{22}& 0 \\\hline
0      &0     & 1 \\\end{array}\right)$}
\rput(190,0){$\vec \xi_{\underline{1}}$}
\rput(165,40){$\vec \xi_{\underline{2}}$}
\rput(190,55){$\vec \xi_3$}
\psline{->}(160,0)(203,10) \pscircle*(160,0){1} \rput(155,0){1} \pscircle*(166,60){1} \rput(161,60){2}
\psline{->}(178,4)(166,60) \pscircle*(203,10){1} \rput(203,5){3}
\psline{->}(173.5,24.5)(203,60)\pscircle*(203,60){1} \rput(203,65){4}
\end{picture}\end{center}
\caption{Initial coordinate system (left), suitable for calculating the interaction between 1 and 2, rotated coordinate system (right), suitable for calculating the interaction between 1 and 3 and the matrix of kinematic rotation that transforms the coordinates $\vec \xi_i $ into $\vec \xi_{\underline{i}} $.}
\label{4part}
\end{figure}

 Now we turn to the core part of the paper, the determination of the matrix elements of the two-body potentials between each pair of particles, $V_{ij}(\mid \vec  r_{ij} \mid)$. 
Clearly the choice of Jacobi coordinates operated above has the virtue of separating the center of mass motion, but 
admittedly complicates the calculation of the matrix elements we are after, except $V_{12}$, because, by construction $\vec \xi_1 \equiv  \sqrt{\mu/\mu_1}\vec r_{12}$. Notice that the 'first coordinates pair', namely $12$, has this special property, while all other pairs ${ij}$ don't, therefore the calculation of matrix elements of $V_{ij}(\mid \vec r_i - \vec r_j \mid)$ is not separable. This is a long-standing problem of which we offer a solution by merging group theoretical and analytic techniques.

i) First, we have to recall the concept of kinematic rotations, also called democracy transformations \cite{LoGa,BaNo,Fano}.
These are linear transformations between different choices (i.e. orderings) of Jacobi coordinates, viz.
\be
\vec \xi_{\un{i}} = \sum_{j=1}^{N-1} D_{ij} \vec \xi_j \;,
\label{rot} 
\ee
where underlined denotes transformed coordinates.
The set of matrices $\mathbb{D}$  with matrix elements $D_{ij}$ form the democracy group that includes rotations and reflections. As these transformations are norm-conserving, the group can be identified with the orthogonal group O(N-1). 
 In the Jacobi coordinates system defined above, the calculation of matrix elements of $V_{ij}(\mid \vec  r_{ij} \mid)$ involves exactly the set of the first $j$ coordinates, namely  $\{ \vec \xi_{1}, \dots, \vec \xi_{j} \}$. 
We call these {\it hot  coordinates} and the others {\it cold  coordinates}.
For example, in the case of four particles depicted in Fig. \ref{4part}, the calculation of the matrix element of $V_{13}$ involves the 6-dimensional integral 
$\langle \phi(\vec \xi_1)\phi(\vec \xi_2)\mid V_{13}(\mid \vec \xi_2- \vec \xi_1/2\mid )\mid \phi(\vec \xi_1)\phi(\vec \xi_2)\rangle$, while the third coordinate can be separated out. In order to achieve a full separation, it is necessary to perform a kinematic rotation in the subspace of hot coordinates that brings one of them to $\vec r_{ij}$ .
This can be viewed as a block-matrix of the form:
\be
\left( 
\begin{array}{c}
\red \vec \xi _{\un{1}}  \\ 
\red \vdots \\
\red \vec \xi _{\un{j}}  \\ 
 \vec \xi _{\un{j+1}}  \\ 
\vdots \\
\vec \xi _{\un{N-1}}  \\
\end{array} \right)
=
\left( 
\begin{array}{c|c}
& \\
\red ~~\mathbb{D}_{j\times j}~~&   \mathbbm{O}\\ 
& \\
& \\ \hline
& \\
 \mathbbm{O}&  \mathbbm{1}_{N-1-j} \\
&  \\
\end{array} \right)
\left( 
\begin{array}{c}
\red \vec \xi _{1}  \\
\red \vdots \\
\red \vec \xi _{j}  \\
\vec \xi _{j+1}  \\
\vdots \\
\vec \xi _{N-1}  \\
\end{array} \right)
\ee
where $\mathbbm{1}$ and $\mathbbm{O}$ are the identity and null (sub-)matrices of appropriate dimensions and the hot coordinates have been highlighted.
We can always choose the new set of hot coordinates in such a way that one of them coincides with $\vec r_{ij}$ and the others connect (sub-)clusters of increasing size as usual (since the order is immaterial to the final result, we choose the first hot coordinate, i.e. the $\un{1}$-th as $\vec  r_{ij}$ coordinate). The kinematic rotation proposed here allows to factorize the matrix element of $V_{ij}$ into the matrix element acting only on bra's and ket's of a single relevant coordinate $\vec \xi_{\un{1}}$ and a product of $(j-1)$ Kronecker delta's, {\it thus simplifying noticeably the problem.}

In Fig. \ref{4part} we give an example for the case of 4 particles. The initial coordinates $\vec \xi_i $, that are suitable for calculating the matrix elements of the $V_{12}$ interaction, are rotated into $\vec \xi_{\underline{i}}$, that are suitable for calculating the matrix elements of $V_{13}$. Notice that the third coordinate (that is a cold coordinate here) is not rotated. If we call $U_{ord}$ the $N$-dimensional nunitary transformation (\ref{tr}) that brings the initial coordinates of $N$ particles to a certain set of mass-scaled Jacobi coordinates with a given  of the labels, then the democracy transformation $\mathbb{D}$ between any two such ordering is given by the $(N-1)$-dimensional submatrix of the product $U_{ord}(U_{ord'})^{-1}$. For the sake of simplicity, we will be concerned only with orderings that brings the relevant pair ${ij}$ to the 'first' position.

ii) Now that we have specified how to change coordinates, before proceeding to the calculation of matrix elements, we have to find the relation between the original and transformed basis states. This crucial step is not easily accomplished in general, but for harmonic oscillator bases one can derive analytic formulas.
The idea is that we need to expand each basis substate in terms of the basis states of the transformed coordinates as follows 
(keeping in mind that the coordinates $j+1,\cdots, N-1$ are not rotated):
\bea
\mid \phi^{1}_{\{\nu_{1}\}}(\vec \xi_{1}) \times \dots \times\phi^{N-1}_{\{\nu_{N-1}\}}(\vec \xi_{N-1}) \rangle = \nonumber \\
 \nonumber \\
=\sum_{\{\un{c}\}} \sigma_c^{\un{c}} \mid \phi^{\un{1}}_{\{\nu_{\un{1}}\}}(\vec \xi_{\un{1}}) \times \dots \times \phi^{\un{N-1}}_{\{\nu_{\un{N-1}}\}}(\vec \xi_{\un{N-1}}) \rangle \;,
\label{kr}
\eea
 where $\sigma_c^{\un{c}} $ are expansion coefficients labeled by the set of all quantum numbers needed in the transformed basis, i.e. $\un{c} =\{\nu_{\un{1}}, \dots, \nu_{\un{N-1}} \}$. 
From Eq. (\ref{kr}), one gets 
\be
\sigma_c^{\un{c}} =  \langle \phi^{\un{1}}_{\{\nu_{\un{1}}\}}(\vec \xi_{\un{1}}) \times \dots \times \phi^{\un{N-1}}_{\{\nu_{\un{N-1}}\}}(\vec \xi_{\un{N-1}}) \mid \phi^{1}_{\{\nu_{1}\}}(\vec \xi_{1}) \times \dots \times\phi^{N-1}_{\{\nu_{N-1}\}}(\vec \xi_{N-1}) \rangle  \;,
\ee
that, in principle, cannot factorize into a product of $N-1$ overlaps, because each rotated $\vec \xi_{\un{k}}$ depends on all the non-rotated $\vec \xi_{k}$, according to Eq. (\ref{rot}). This means that the integral would be an almost inextricable $3j$-dimensional tangle for large values of $j$. Fortunately, at least in the case of a set of $(N-1)$ decoupled harmonic oscillator bases, we can solve this problem analytically. {\it The calculation of the $\sigma$ coefficients can be recast in terms of summations of certain coefficients, that come from the repeated application of the umbral identities for Hermite polynomials and from the series definition of the latter, and of Gamma functions, that come from the analytic integration of the remaining rotated polynomials}. 
The lengthy derivation and generalization will be described elsewhere. Of course the numerical evaluation of these analytic expressions is hundreds of times quicker than the brute-force calculation of multidimensional integrals and this is a considerable advantage.

Using expansion (\ref{kr}), the matrix elements that we are seeking are therefore given by

\bea
\langle \tilde \Phi^{(N)}_{c'} \mid V_{ij}(\mid \vec r_{ij} \mid) \mid  \tilde \Phi^{(N)}_c \rangle = 
\delta_{\{\nu_{j+1}'\}\{\nu_{j+1}\}} \dots \delta_{\{\nu_{N-1}'\}\{\nu_{N-1}\}}  \nonumber \\
\sum_{\un{c'}, \un{c}} {\sigma_{c'}^{\un{c'}}}^*  ~\sigma_{c}^{\un{c}}  
\langle \phi^{\un{1}}_{\nu_{\un{1}}'}(\vec \xi_{\un{1}}) \mid V_{ij}(\xi_{\un{1}}) \mid  \phi^{\un{1}}_{\nu_{\un{1}}}(\vec \xi_{\un{1}})  \rangle  
\delta_{\{\nu_{\un{2}}'\}\{\nu_{\un{2}}\}} \dots \delta_{\{\nu_{\un{j}}'\}\{\nu_{\un{j}}\}}  \;,
\label{mev}
\eea
where some care must be taken with the labeling of indexes of $\nu$ (unprimed means ket, primed means bra, underlined means transformed).
The last matrix element can be reduced to the one-dimensional integral in the radial variable $\langle \phi^{\un{i}}_{\nu_{\un{i}}'}(\xi_{\un{i}}) \mid V_{ij} \mid  \phi^{\un{i}}_{\nu_{\un{i}}}(\xi_{\un{i}})  \rangle $ for central potentials. Notice, in fact, that the coefficients $\sigma$ of Eq.(\ref{mev})  differ only in the last set of quantum numbers (associated with the $\un{1}$ coordinate) that can be further reduced to just the two principal quantum numbers $n_{\un{1}}'$ and $n_{\un{1}}$ , respectively.

Eqs. (\ref{met}), (\ref{meu})  and (\ref{mev})  give a {\it new method  to diagonalize the N-body Schr\"odin\-ger equation with two-body interactions}. This method proceeds through exact analytical steps, albeit in practice one must decide a truncation in the number of quanta for each $\phi$ in Eq.(\ref{expa}). Numerical approximations come into play only in the calculation of one dimensional integrals and in the matrix diagonalization. Several exact formulas exist for certain analytic functions of the distance (take for example integrals of powers in the h.o.), therefore the calculation of matrix elements can often be done exactly.
In many cases, after diagonalization a proper symmetrization procedure must be adopted. This method transfers the (often untreatable) complexity of the calculation of matrix elements of the mutual particle-particle interaction to a number of simple overlaps and well-known matrix elements of a single coordinate, thereby allowing a straightforward solution of the formidable many-body problem.

\section{Validation in the 1D three-body case}
Before embarking on a longer campaign of theoretical studies involving more sophisticated calculations, we need demonstrate the validity of our method on analytic cases. There are only a handful solved models \cite{Cal1, Cal2,Suth}, among which we choose the well-known one dimensional Calogero model with harmonic pairwise interactions of the type $V_{ij}=\hbar \omega (x_i-x_j)^2$. The exact spectrum for N=3 particles with Bose statistics on a line is found in Ref. \cite{Cal2}, namely $E_{Cal} = \sqrt{3}\hbar\omega (2n+l+1)$ with $n,l$ non-negative integers, and reproducing this result it's easy with our method. Essentially, we just use the matrix elements of $x^2$ between harmonic oscillator basis states and use the exact formula for the $\sigma$ coefficients in the calculations of $V_{13}$ and $V_{23}$. 
We give here only the final result for three particles in 1D, where the sets of quantum numbers simply reduces to the number of oscillator quanta:
\be
\sigma_c^{\un{c}} = \frac{2^{-(n_1+n_2+n_{\un{1}}+n_{\un{2}})/2}}{ \pi\sqrt{n_1!n_2!n_{\un{1}}!n_{\un{2}}!}}
\sum_{k=0}^{n_1} {n_1 \choose k}\sum_{j=0}^{n_2}{n_2\choose j} {\cal I}(n_1,n_{\un{1}},k,j,\mathbb{D}){\cal I}(n_2,n_{\un{2}},k,j,\mathbb{D})
\ee
where ${\cal I}$ are some one-dimensional integrals that can be written in terms of summations of Gamma functions and $\mathbb{D}$ is the democracy transformation between the set 'starting' with 12 and the one with either 13 or 23 as 'first coordinate'.

\begin{figure}[!t]
\begin{center}
\begin{picture}(140,85)
\psset{unit=1pt}
\psline{->}(0,0)(0,80)\rput(5,75){$E$}
\psline{->}(0,0)(120,0) \rput(110,5){$(n,l)$}
\psline{-}(10,10)(30,10)
\rput(20,5){(0,0)}
\psline{-}(30,30)(50,30)
\rput(40,25){(0,1)}
\psline{-}(10,50)(30,50)\psline{-}(50,50)(70,50)
\rput(20,45){(1,0)}\rput(60,45){(0,2)}
\psline{-}(30,70)(50,70)\psline{-}(70,70)(90,70)
\rput(40,65){(1,1)}\rput(80,65){(0,3)}
\rput(135,85){deg.}
\rput(135,65){2}\rput(135,45){2}\rput(135,25){1}\rput(135,5){1}
\end{picture}\end{center}
\caption{Spectrum of the Calogero linear model (1D) with harmonic pairwise interactions. The analytic result in terms of quantum numbers $(n,l)$ is obtained easily with our numerical method. Degeneration of energy levels is shown on the right.}
\label{calo}
\end{figure}
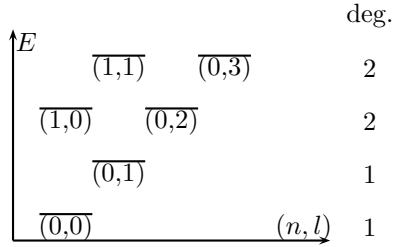

In a basis truncated at $N_q=15$ quanta, we get the linear spectrum of Calogero effortlessly in the turn of a few seconds on a table-top machine, with a numerical precision that is sufficient to all practical purposes and with correct degeneration of energy levels. The lowest eigenvalues are shown in Fig. \ref{calo}.

In our opinion, the mathematical derivation of the new method and the proof-of-principle discussed in the present paper hold great promises and pave the way to studies of greater impact, because we have now a handy theoretical tool that allows to explore several physics systems: from academic cases such as the spectrum and wavefunctions of $N$ particles in 1D, to realistic models of few-body systems (atoms, molecules, nuclei, BEC) in 3D and to more advanced ideas such as for instance Efimov states, condensation, exotic systems, phase transitions in stable and unstable systems, etc.

\section*{Acknowledgements}
{\it In:Theory}, PRAT Project n. CPDA154713, Univ. Padova (Italy).

L.F. acknowledges fruitful discussions with A.Richter and V.Efros (in 2011) and with A.Vitturi (2011-2016). Most of the present material has been conceived while working at the ECT*(Trento) in 2011, but could only be finalized now. 

The computing facilities offered by CloudVeneto (CSIA Padova and INFN) are acknowledged.

\end{document}